\documentclass[12pt,a4paper]{article}
\usepackage[latin1]{inputenc}
\usepackage[T1]{fontenc}
\usepackage{amsmath}
\usepackage{amsfonts}
\usepackage{amssymb}
\usepackage[english]{babel}
\usepackage[dvips]{graphicx,color}
\newcommand{\be}{\begin{equation}}
\newcommand{\ee}{\end{equation}}

\newcommand{\fii}{\varphi}
\newcommand{\fiid}{\varphi^*}
\newcommand{\ecua}{\begin{equation}}
\newcommand{\fin}{\end{equation}}
\newcommand{\punto}{\;\; .}
\newcommand{\coma}{\;\; ,}

{\catcode`\|=\active\gdef\Braket#1{\left<\mathcode`\|"8000\let|\bravert {#1}\right>}} 
\def\bravert{\egroup\,\vrule\,\bgroup}
\title{Casimir effect in 2+1 dimensional noncommutative theories}
\author{C.~D.~Fosco
and
G.~A.~Moreno\\
{\normalsize\it Centro At\'omico Bariloche and Instituto Balseiro}\\
{\normalsize\it Comisi\'on Nacional de Energ\'\i a At\'omica}
\\
{\normalsize\it R8402AGP Bariloche, Argentina.}}
\begin{document}
\date{\today}
\maketitle
\begin{abstract}
We study the Dirichlet Casimir effect for a complex scalar field on two
noncommutative spatial coordinates plus a commutative time.  To that end,
we introduce Dirichlet-like boundary conditions on a curve contained in the
spatial plane, in such a way that the correct commutative limit can be
reached. We evaluate the resulting Casimir energy for two different curves:
(a) Two parallel lines separated by a distance $L$, and (b) a circle of
radius $R$.  In the first case, the resulting Casimir energy agrees exactly
with the one corresponding to the commutative case, regardless of the
values of $L$ and of the noncommutativity scale $\theta$, while for the
latter the commutative behaviour is only recovered when \mbox{$R >>
\sqrt{\theta}$}. Outside of that regime, the dependence of the energy with
$R$ is substantially changed due to noncommutative corrections, becoming
regular for $R \to 0$.  
\end{abstract}
\newpage
In the Casimir effect~\cite{intro}, a nice interplay between the geometry of a spatial
region and the vacuum fluctuations of a field conspire to produce an
observable effect: the Casimir force. The properties of such a force do
depend on the kind of field theory considered, on the nature of the
boundary conditions imposed, and on the number of spatial dimensions.  The
physical reason is that the properties above will determine the kind of
vacuum fluctuations that are allowed in each spatial region, and whose
competing effects produce the Casimir force. 

On the other hand, Noncommutative Quantum Field Theories (NCQFT's) ~\cite{Douglas:2001ba}, are
endowed with an intrinsic scale, due to the fundamental
commutation relation:
\begin{equation}
[x_\mu \,,\, x_\nu ] \;=\; i \, \theta_{\mu\nu} \;\;, \;\;\;\mu,\,\nu =
\,0,\,1,\,\ldots,\,d\; \;\;\;,
\end{equation} 
where $\theta_{\mu\nu}$ is a constant antisymmetric tensor.  The resulting
existence of a `granularity' for the coordinates resolution, with its
corresponding scale $\theta$ playing the role of a minimal area, suggests
the possibility that noncommutativity might affect the properties of the
Casimir force introducing corrections depending on $\sqrt{\theta}/L$ (where $L$ is a length related to the `size' of the system).  

Besides this immediate, merely dimensional argument, we should expect also
interesting results to emerge when a NCQFT is subject to boundary
conditions on a non trivial region: firstly, the boundary conditions
are certainly problematic by themselves, since they are imposed
on elements in a noncommutative algebra. In particular, the act of imposing a boundary
condition on a codimension-$1$ manifold will set the spatial resolution 
along one spatial coordinate to zero.  Secondly, NCQFT's have been associated to 
{\em incompressible\/} quantum fluids~\cite{Susskind:2001fb,Jackiw:2004nm}, whose fluctuations are  (because of that property) expected to be more sensitive to the 
existence of boundaries than in the commutative case. 

In this letter, we consider the Casimir effect for the NCQFT of a complex
scalar field in $2+1$ dimensions. In this case only two spacetime
coordinates may be noncommutative; we shall assume them to be the two
spatial ones (which form a Moyal plane), while the time is a commutative
object.  Our main motivation for considering precisely this situation is
that concrete  physical systems do exist where noncommutativity is
naturally realized in exactly that way: indeed, when a strong constant
magnetic field is applied to an essentially two-dimensional system, a
projection to the lowest Landau level justifies a noncommutative
description~\cite{Dunne:1992ew,LLL}. On the other hand, since the time coordinate remains
commutative, the Hamiltonian still plays the role of the generator of time
translations in the usual way, hence many standard Quantum Field Theory
tools have the same interpretation that in the commutative case. In
particular, a path integral formula for the vacuum persistence amplitude
can be applied to obtain the vacuum energy.     

In this way we shall be able to disentangle new effects that result from
the interference of noncommutativity and boundary conditions, from the ones
that, even in the absence of boundaries, could still modify the vacuum
energy.

Some works have already dealt with the issue of imposing boundary
conditions within the context of NCQFT~\cite{casimir_trabajos,casimir_trabajos2,casimir_trabajos3}. 
However, both the kind of system considered and the approach followed are different; therefore the ensuing conclusions are incommensurable. For example, in~\cite{casimir_trabajos}, the time coordinate is regarded as noncommutative, while in~\cite{casimir_trabajos2} and~\cite{casimir_trabajos3} noncommutativity is introduced for manifolds without 
boundaries.

The complex scalar field $\varphi$, on which boundary conditions are to
be imposed on the curve ${\mathcal C}$, shall be
equipped with a standard free Euclidean action $S_0$: 
\begin{equation}\label{eq:defs0}
S_0(\fiid, \fii)=\int d^3x\; \big(\partial_\mu \varphi^* \star 
\partial_\mu \varphi + m^2 \varphi^* \star \varphi \big) \;,
\end{equation} 
where the Moyal product involves just the two spatial coordinates $x_j$,
$j=1,2$: 
\begin{equation}
f(x_0,x_1,x_2) \star g(x_0,x_1,x_2) \;\equiv\; \lim_{y \to x} \,
e^{\frac{i}{2} \theta_{jk}\frac{\partial}{\partial x_j} \frac{\partial}{\partial y_k}} 
f(x_0,x_1,x_2) \, g(x_0,y_1,y_2) \;. 
\end{equation}
To impose the boundary conditions for the field on ${\mathcal C}$, we use 
the procedure of adding to the Lagrangian a term that introduces an 
interaction with ${\mathcal C}$, in such a way that the boundary conditions
emerge when the interaction is strong. This procedure, already used 
in the Commutative Quantum Field Theory (CQFT) case \cite{jaffe}, is here much simpler
than attempting to impose the bondary conditions on the field. 

To briefly review this approach, let as apply it to the commutative version of our first
example, namely, a region ${\mathcal C}$ that corresponds to two straight
lines at $x_2=0$ and $x_2=L$. In this case, the total Euclidean action $S =
S_0 + S_I$ includes an interaction with ${\mathcal C}$:
\begin{equation}\label{comprob} 
S_I(\fii, \fiid)= \lambda \int_{x_0,x_1} 
\big[ \fiid(x_0,x_1,0)\fii(x_0,x_1,0) +\;\fiid(x_0,x_1,L)\fii(x_0,x_1,L) \big]  \;.
\end{equation} 
The vacuum energy $E_0$, may be obtained from the path integral expression
\begin{equation}\label{eq:path1}
e^{- T E_0} \,=\, \frac{\int {\mathcal D} \varphi^* {\mathcal D} \varphi
\; e^{-S}}{\int {\mathcal D} \varphi^* {\mathcal D} \varphi \; e^{-S_0}}
\end{equation}
where the denominator subtracts the $L \to \infty$ contribution, and $T$ is
assumed to tend to infinity. Then 
\begin{equation}\label{basic}
 E_0\;=\; \lim_{T \to \infty, \lambda \to \infty} \, \frac{1}{T} \, {\rm Tr} \log \big( 1+
\Delta D \big)
\end{equation}
where $\Delta$ is the free propagator and $D$ is an operator
whose kernel is defined by
\begin{equation}
S_I=\int_{x,\;y}\fiid(x)\;D(x,y,L,\lambda)\;\fii(y) \punto
\end{equation}
Of course, in this case $E$  is expected to be proportional to the length
of the lines (in the $x_1$ direction). Since that length is regarded as infinite, 
in practice one deals with the linear density of energy. 
The $\lambda \to \infty$ limit is, on the other hand, taken in order to enforce 
Dirichlet boundary conditions.

Let us now generalize this example to the noncommutative case, considering
an action \mbox{$S^\star = S_0 + S_I^\star$} where  
\mbox{$S_I^\star \equiv S_I^{\star (L)}+S_I^{\star (0)}$}
with:
\begin{equation}
S_I^{\star (L)} \;\equiv \; \lambda \,\int_x \fiid \star \delta_2^L \star \fii \coma 
\end{equation}
and $\delta_2^L \equiv \delta{(x_2-L)}$. $S_I^{\star (0)}$ corresponds to
setting $L\equiv 0$ above. The Casimir energy will then be obtained by applying
(\ref{eq:path1}) to the action $S^\star$.

Introducing the Fourier transform of the field with respect to the $x_0$ and $x_1$ variables,
\be
\fii(x_0,x_1,x_2) \,=\, \int \frac{d\omega}{2\pi} \int \frac{dp}{2\pi} \,  
e^{i (\omega \, x_0 \,+\, p \, x_1)} \; \tilde\fii (\omega,p,x_2)
\coma
\ee
and using the properties of the $\star$-product, we may write the
interaction term at $x_2=L$ as follows:
\begin{eqnarray}
S^{\star (L)}_I &=& \lambda  \int \frac{d\omega}{2\pi} \int \frac{dp}{2\pi} \int dx_2 
\, \delta(x_2-L) \nonumber\\
&\times& \tilde\fiid (\omega,\; p,\; x_2+\frac{\theta p}{2})\;
\tilde\fii(\omega,\; p,\; x_2+\frac{\theta p}{2}) \;,
\end{eqnarray}
and a similar expression for $S^{\star (0)}_I$.  
Performing now the change of variables 
\begin{equation}
\tilde\varphi(\omega,\;p,\;x_2) = \psi(\omega,p,x_2-\frac{\theta p}{2})
\end{equation}
(which yields no Jacobian in the path integral), and taking into account the invariance of the
free kernel under translations in $x_2$, one sees that the action becomes: 
$$
S \;=\; \int \frac{d\omega}{2\pi} \int \frac{dp}{2\pi} \int dx_2 \;
\psi^*(\omega, p, x_2) \; \Big\{ 
(\omega^2+p^2-\partial_2^{2}+m^2)
$$
\begin{equation}
+\; \lambda \big[ \delta(x_2-L) + \delta(x_2) \big] \Big\} \,  
\psi(\omega,\;p,\;x_2)
\punto
\end{equation}
Note that $\theta$ has disappeared from the action, and indeed, this
expression coincides with the one we would have obtained in the commutative
case. This means that the vacuum energy $E_0$ for the noncommutative model
is identical to the commutative one, regardless of the value of $\lambda$.
In particular, for the Dirichlet case ($\lambda \to \infty$), we conclude
that the Casimir force in the NCQFT agrees, for this geometry, with the
CQFT one.

This property could seem to be surprising at first, but then one should
realize that it is a consequence of the fact that this boundary divides
space into two noncompact subsets. And the noncommutative effects seem to
be controlled by the ratio between the area enclosed by the boundary and
the minimal area $\theta$.
It should be noted that the agreement with the CQFT result is realized
after performing a field redefinition that depends on $\theta$. This means
that the $\langle \varphi \varphi^*\rangle$ propagator in the presence of
the boundaries will not be equal to its commutative counterpart, in spite
of the fact that they will produce the same result for the Casimir energy.

Let us now consider the qualitatively different case of a circular defect;
to be more precise, assuming a free action as before, we now 
consider the NCQFT analog of a commutative interaction term: 
\be
S_I\;=\;\lambda\;\int d^3 x \, \delta(r-R)\;\fiid \fii \;, 
\ee 
in the $\lambda \to \infty$ limit. 
A difficulty one immediately faces is to find a natural way to introduce
the noncommutative version of the $\delta(r-R)$-function.  However, that is
not strictly necessary: we only need to assign a meaning to the integral of
$\delta(r-R)$ times a function (as it appears in the interaction term).
From the defining properties of the $\delta$ distribution in the
commutative case, we recall that it only depends on the values of the
function on the $x_1^2 + x_2^2 = R^2$ circle.  And there is a basis for the space of fields
where this problem looks somewhat simpler, since it is compatible with
rotation symmetry in the noncommutative plane: the so called `matrix
basis' \cite{gracia}. Here, functions that depend only on $R$ are diagonal, and one can
then attribute a clear meaning to the interaction term, as one that only
depends on the value of the field on an eigenspace of $x_1^2 + x_2^2$.

Using the same conventions as in~\cite{oneloop}, we shall assume the
interaction term to have the form: 
\begin{equation}\label{ec:terminoanillo}
S_I^\star \;=\; \lambda \int d^3x 
\;f_{NN}\star \fiid \star f_{NN} \star \fii  \;,
\end{equation}
where no sum over $N$ is meant.
As it has been shown in~\cite{oneloop}, $f_{NN}$ is a radial function. 
 And certainly it yields for the interaction term a result that only
depends on the function at a radius which is determined by $N$:  
recalling the relation $x^2+y^2 \; \longleftrightarrow \;
2(N+\frac{1}{2})\;\theta$, $R\;\approx\;\sqrt{2N\theta\;}$, which becomes a
continuous variable in the commutative limit (large $N$). For small $N$,
only discrete values of $R$ are possible: as expected, there is an `area
quantization' effect and one cannot confine the field to a region that whose area is not a multiple of the minimal one.
The commutative Casimir energy for this case behaves like $R^{-1}$, which in our case 
would correspond to $N^{-\frac{1}{2}}$. 

Decomposing the field variables in the matrix base, 
\mbox{$S_I^\star = \lambda \int _{x_0}\, \fiid_{NN}(t) \fii_{NN}(t)$} \footnote{Global factors are absorbed in a redefinition of the field variable because the action is exactly quadratic.}. 
Then, the vacuum energy becomes:
\begin{equation}\label{ec:ppal_e}
E_0(N) \;=\;\int \frac{d\omega}{2\pi} \log \big(1+\lambda \; \Delta_{N,N;N,N}(\omega)\big)
\;,
\end{equation}
where $\Delta_{n_1,n_2,n_3,n_4}$ is the free propagator written in the
matrix basis. We may obtain it by a simple redefinition from the $1+1$
dimensional one presented explicitly in~\cite{grosse-wullkenhar}, the
result being:
\be
\Delta_{n_1n_2,n_3n_4}(\omega)=\delta_{n_1+n_3,n_2+n_4}\int_{0}^{\infty}dx\, 
x^{n_2-n_1} \; e^{-x} \; \sqrt{\frac{n_1! n_4!}{n_2! n_3!}} \;
\frac{\mathbb{L}^{n_2-n_1}_{n_1}(x) \;
\mathbb{L}^{n_3-n_4}_{n_4}(x)}{\omega^2+m^2+\frac{2}{\theta}x} \;,
\ee
where the $\mathbb{L}^a_b$ denote associated Laguerre polynomials.
In our case only part of the diagonal elements of this object appear, so
that the expression for the vacuum energy becomes:
\be \label{ec:ppal_posta} 
E_0(N)=\frac{1}{2\pi}\int_{\omega \in
\mathbb{R}}\log \Big{(}1+\lambda \int_{0}^{\infty} dx \; e^{-x}\;
\frac{[\mathbb{L}^N(x)]^2}{\omega^2+m^2+\frac{2}{\theta} x}\Big{)} \;,
\ee
where $\mathbb{L}^N$ is the Laguerre polynomial of order $N$.

The previous result for the vacuum energy is the starting point for our
derivation of more explicit expressions, in different limits and for
particular cases. 

We first assume $m=0$; thus, changing variables: $\omega \to \theta^{-1/2}
\omega$, we have:
\be
E_0(N)=\frac{1}{2\pi \sqrt{\theta}}\int_{\omega \in \mathbb{R}}\log
\Big{(}1+\lambda \theta \int_{0}^{\infty} dx\; e^{-x}\;
\frac{[\mathbb{L}^N(x)]^2}{\omega^2+2 x}\;\Big{)} \punto
\ee
If the condition $\lambda \theta << 1$ is met, we have:
\begin{eqnarray} 
E_0(N) &\approx& 
\frac{1}{2\pi \sqrt{\theta}}\; \lambda \theta \;
\int_{\omega \in \mathbb{R}}\int_{0}^{\infty}dx \; 
e^{-x}\; \frac{[\mathbb{L}^N(x)]^2}{\omega^2+2 x} 
\nonumber\\
&=&\frac{1}{2\pi \sqrt{\theta}}\; \lambda \theta \pi \;
\int_{0}^{\infty}dx \; e^{-x}\; \frac{[\mathbb{L}^N(x)]^2}{\sqrt{2 x}} 
\punto \label{aprox1}
\end{eqnarray}

\begin{figure}[!h] 
\begin{center}
\includegraphics[angle=270,width=13cm]{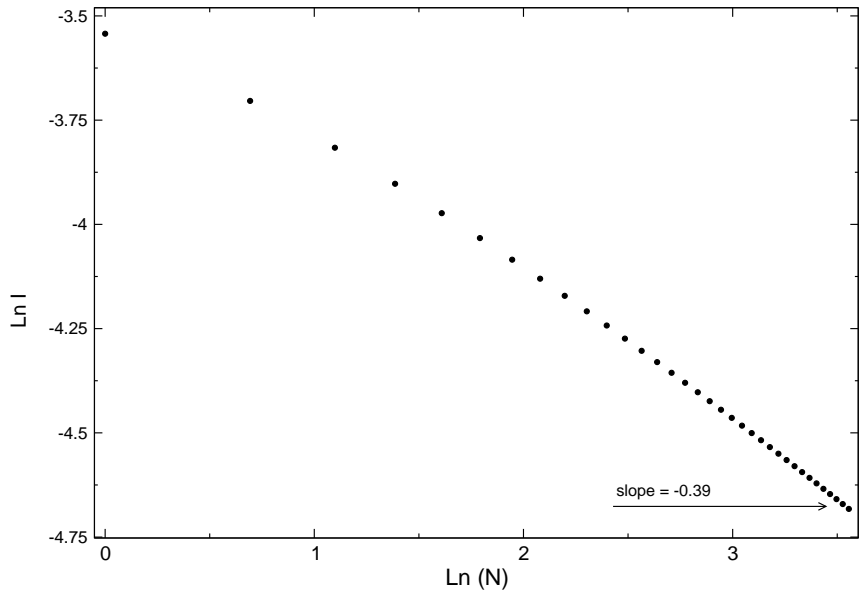}%
\end{center}
\caption{Numerical evaluation of the integral in (\ref{aprox1}).} \label{fi:numerico_integral1}
\end{figure}

Of course this is a convergent integral. We performed a numerical evaluation of (\ref{aprox1}) for different values of $N$,  the
results of which are shown in Figure
\ref{fi:numerico_integral1}.  
Note that close to the origin $E_0$ is well behaved;
for large $N$ we should have instead an asymptotic behaviour \mbox{$\sim
1/\sqrt{N}$}.

\begin{figure}[!h] 
\begin{center}
\includegraphics[angle=270,width=13cm]{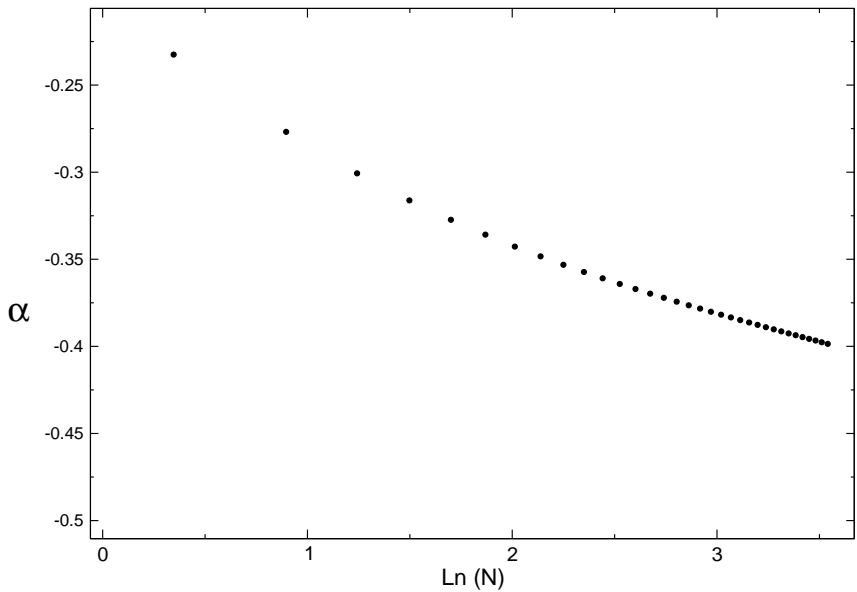}%
\end{center}
\caption{Slope ($\alpha$) of Figure \ref{fi:numerico_integral1}.} \label{fi:slope}
\end{figure}

\begin{figure}[!h] 
\begin{center}
\includegraphics[angle=270,width=13cm]{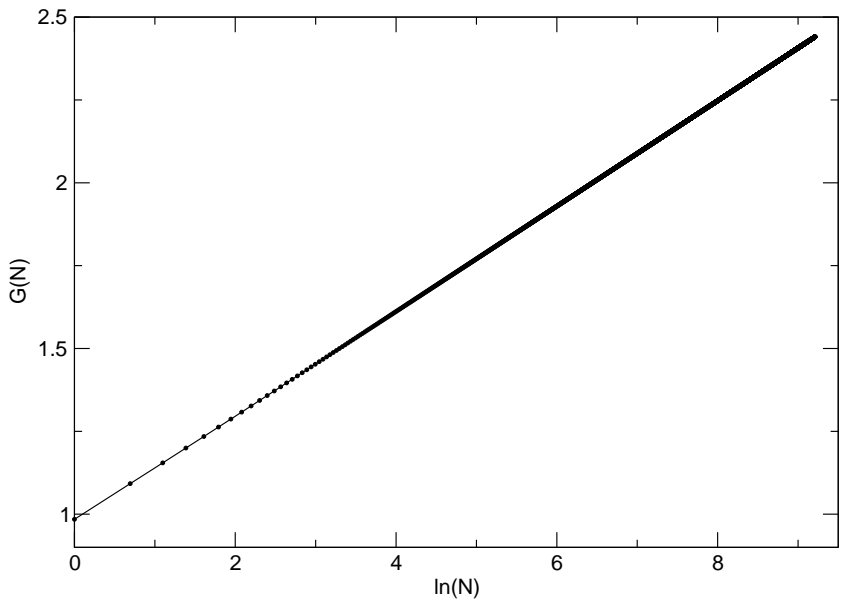}%
\end{center}
\caption{Numerical evaluation of the function $G(N)$.} 
\label{fi:loggn}
\end{figure}

We see that, up to our maximum $N$, (\ref{aprox1}) does not yet reach its
asymptotic regime, which corresponds to a $-\frac{1}{2}$ slope~\footnote{We have defined
the coefficient of the power law as the one an experimentalist would
use, namely, $\alpha=\frac{\partial \log (\Delta E)}{\partial \log
(N)}$.}. In Figure \ref{fi:slope} we plot the slope ($\alpha$) of the previous graph versus 
$\log N$ .
\\
To see that the asymptotic power law will be such that $\alpha \rightarrow -\frac{1}{2}$,  we step back to:
\be \label{bak}
E_0(N)=\frac{1}{2\pi \sqrt{\theta}}
\int_{\omega \in \mathbb{R}}\log \Big{(}1+\lambda \theta \int_{0}^{\infty} \; e^{-x}\; \frac{[\mathbb{L}^N(x)]^2}{\omega^2+2 x}\; dx\Big{)} 
\punto
\ee
In the large-$N$ limit, we may use have the property:
\be
\lim_{n\rightarrow \infty} \mathbb{L}^n\big{(}\frac{z^2}{4n}\big{)}=J_0(z)
\coma
\ee
where $J_0$ is the Bessel $J$ function of order zero. This approximation can be used inside the integral in (\ref{bak}), because 
the pre-factor reduces the effective domain of integration. Thus, 
if $N>>1$:
\be
\int_{0}^{\infty}dx\, \frac{e^{-x}\;[\mathbb{L}^N(x)]^2}{\omega^2+2x}\;
\approx \frac{1}{2N}\int_{0}^{\infty}dz z 
\frac{e^{-\frac{z^2}{4N}}\;(J_0(z))^2}{\omega^2+\frac{2z^2}{4N}}
\punto
\ee
To proceed, we only need $\lambda \theta $ to be bounded, so that for a
large enough $N$ ($\lambda \theta /N <<1$), we shall have:
\be
E_0(N) \approx \frac{1}{2\pi \sqrt{\theta}}\int_{\omega \in
\mathbb{R}}\frac{\lambda
\theta}{2N}\int_{0}^{\infty}\frac{e^{-\frac{z^2}{4N}}\;(J_0(z))^2}{\omega^2+\frac{z^2}{2N}}\;zdz
\coma
\ee
or:
\be
\Delta E \approx \frac{\lambda \theta}{2\sqrt{2\theta N}} \int_{0}^{\infty}\;\big{(}J_0(z)\big{)}^2\;e^{\frac{-z^2}{4N}}\;dz 
\punto
\ee
In this manner we have managed to extract a $\frac{1}{\sqrt{2\theta N}}$ dependence, 
but we still have to deal with the function:
 \be
G(N)\;=\;\int_{0}^{\infty}(J_0(z))^2\;e^{-\frac{z^2}{4N}}\;dz
\punto
\ee
A numerical study of this function shows that it diverges logarithmically (a plot is shown in Figure \ref{fi:loggn}), so the
asymptotic power law holds, as we have claimed before~\footnote{Using our definition of $\alpha$ we find : $\alpha = -\;\frac{1}{2}+\frac{B}{\underbrace{A+B\log (N)}_{G(N)}}$}.
We thus see  the result converges to the asymptotic regime slowly, since the 
limit is approached logarithmically. Writing $1/N^{\frac{1}{2}+\epsilon}$,
we have, for example,  $\epsilon(N=8000) \simeq \; -0.07$. 
In spite of the slow convergence, the  $1/N^{\frac{1}{2}}$ power law is
indeed asymptotically reached in the large $N$ regime. 

We now show more explicitly that the interaction term (\ref{ec:terminoanillo}) is 
delta-like in the commutative limit. In order to do that, consider the Fourier transform of $f_{NN}$~\cite{oneloop}.
\be
\frac{\hat f_{NN}(k)}{\sqrt{\theta}}=2 \pi \sqrt{\theta}\;e^{-\frac{\theta\;k^2}{4}}\;\mathbb{L}^{(N)}\big{(}\frac{2 N \theta k^2}{4N}\big{)}\;\approx\;2\pi \sqrt{\theta}\;e^{-\frac{\theta k^2}{4}}\;J_0(\sqrt{2N\theta}\;k)
\punto
\ee
So that the inverse reads:
\be
\frac{f_{NN}(r)}{\sqrt{\theta}}=\int_{0}^{2\pi}\int_{0}^{\infty}\frac{\sqrt{\theta}}{2\pi}\;e^{-\frac{\theta k^2}{4}}\;J_0(Rk)\;e^{ikr\cos \beta}\;kdk\;d\beta
\coma
\ee
which using the integral representation of $J_0$ gives
\be
\frac{f_{NN}(r)}{\sqrt{\theta}}=\int_{0}^{\infty}\sqrt{\theta}\;e^{-\frac{\theta k^2}{4}}\;J_0(Rk)\;J_0(kr)\;kdk
\punto
\ee
So, in the limit $\theta \rightarrow 0$, orthogonality relation
\be \nonumber
\int_{0}^{\infty}\;J_{\alpha}(x\;v)\;\;J_{\alpha}(x\;u)\;xdx=\frac{1}{u}\;\delta(u-v) \coma
\ee
yield to  
\be
\frac{f_{NN}(r)}{\sqrt{\theta}} \approx \frac{\sqrt{\theta}}{R}\;\delta(r-R)
\punto
\ee
On the other hand, because of the previous relation, for the second  $\delta$-like factor 
we have the correspondence
\be
\delta(0) \leftrightarrow \frac{1}{R}\;A(\frac{R^2}{\theta})
\coma
\ee
where $A(q)$ is given by
\be \label{ec:aden}
A(q)=\int_{0}^{\infty}\;(J_0(x))^2\;e^{-\frac{x^2}{4q}}\;xdx
\punto
\ee
We have seen numerically that $A(q)\approx 1.77 \sqrt{q}$, thus $\delta(0)\approx \;\frac{1}{\sqrt{\theta}}$.
We the see that the interaction term is
\be \label{ec:terminofinal}
\frac{\sqrt{\theta}}{R^2} \; \lambda \theta \;\; \int\; \delta_R \fiid \fii
\coma
\ee
that, using the assumptions
\begin{displaymath} \left\{
    \begin{array}{ll} 
\lambda \theta\; <<\; N
      \\ 
R^2\;\approx\;2N\theta
    \end{array}\right.
\end{displaymath}
the asymptotic form of the noncommutative interaction term could be rewritten as
\be \nonumber
S^\star_I \;\sim \; g(\theta) \; \xi \; \int\;d^3x \, \delta_R \;\fiid \fii
\;,
\ee
where $g(\theta) \equiv \frac{1}{2\sqrt{\theta}}$ is a large constant with dimension of mass, while $\xi \equiv \frac{\lambda \theta}{N}$ is a small (and can be assumed to be fixed) dimensionless constant. This produces then the `hard' $\delta$-like form in the asymptotic regime,  as claimed at the beginning.

We have seen that, since the defect encloses a bounded region, the vacuum energy shift 
is seriously modified with respect to the commutative case. 
In particular, close to zero size the energy is finite, what can be shown without 
resorting to any approximation.

Noncommutativity effects on the energy extend to large distances, as it was shown the
correction to commutative exponent for the power law $\epsilon$ goes to
zero as $1/\log(N)$.

We have also studied the finite-mass case, where we found that the commutative power law is reached at shorter distances. The asymptotic behavior was studied numerically from expression:
\be\label{ec:conmasamu}
\Delta
E=\frac{\lambda
\theta}{4N\sqrt{\theta}}\int_{0}^{\infty}\frac{z\;dz}{\sqrt{\mu^2+\frac{2z^2}{4N}}}\;e^{-\frac{z^2}{4N}}\;\big{(}J_0(z)\big{)}^2
\coma 
\ee 
where $\mu^2=\theta m^2$, which was deduced from (\ref{ec:ppal_posta}).

\section*{Acknowledgements}
C.\ D.\ F.\ acknowledges support from CONICET, ANPCyT and UNCuyo (Argentina). G.~A.~M. is supported by a Petroenergy SA - Trafigura studentship at Instituto Balseiro UNCuyo.


\end{document}